\documentclass[aps,twocolumn,amsmath,amssymb,footinbib]{revtex4}
\usepackage[colorlinks=true, pdfstartview=FitV, linkcolor=red, citecolor=blue, urlcolor=cyan]{hyperref}
\newcommand{\boldl}{{\it L}}

\def\anp#1#2#3{Annals Phys. {\bf #1}, #2 (#3)}
\def\atmp#1#2#3{Adv. Theor. Math. Phys. {\bf #1}, #2 (#3)}

\def\ibid#1#2#3{{\it ibid.} {\bf #1}, #2 (#3)}

\def\jhep#1#2#3{J. High Energy Phys. #2 (#3) #1}

\def\npa#1#2#3{Nucl. Phys. A {\bf #1}, #2 (#3)}
\def\npb#1#2#3{Nucl. Phys. B {\bf #1}, #2 (#3)}

\def\plb#1#2#3{Phys. Lett. B {\bf #1}, #2 (#3)}

\def\prd#1#2#3{Phys. Rev. D {\bf #1}, #2 (#3)}
\def\prl#1#2#3{Phys. Rev. Lett. {\bf #1}, #2 (#3)}
\def\phr#1#2#3{Phys. Rep. {\bf #1}, #2 (#3)}

\def\rpp#1#2#3{Rep. Prog. Phys. {\bf #1}, #2 (#3)}
\def\rmp#1#2#3{Rev. Mod. Phys. {\bf #1}, #2 (#3)}

\begin{document}
\title{Effective Theory of Wilson Lines and Deconfinement}
\author{Robert D. Pisarski}
\affiliation{
Department of Physics,
Brookhaven National Laboratory, Upton, NY, 11973, U.S.A.\\
}
\begin{abstract}
To study the deconfining phase transition at nonzero temperature,
I outline the perturbative construction of
an effective theory for straight, thermal Wilson lines.
Certain large, time dependent gauge transformations
play a central role.  They imply the existence of interfaces,
which can be used to determine the form of the effective theory as
a gauged, nonlinear sigma model of adjoint matrices.
Especially near the transition, the Wilson line may undergo a Higgs effect.
As an adjoint field, this can generate eigenvalue repulsion
in the effective theory.
\end{abstract}
\date{\today}
\maketitle

Recent results at the Relativistic Heavy Ion Collider (RHIC) demonstrate
qualitatively new behavior for the collisions of heavy ions at high
energies \cite{whitepaper}.  RHIC appears
to have entered a region above $T_c$, the temperature for deconfinement,
reaching up to temperatures a few times $T_c$.
The experimental results cannot be explained if the
transition is directly from a confined phase to a perturbative
Quark-Gluon Plasma (QGP).  Instead, RHIC seems to probe a novel region,
which has been dubbed the ``sQGP'' \cite{sQGP}.

In this paper I sketch how to develop an effective theory for the sQGP.
Classically, the model is a familiar spin system, a gauged
principal chiral field \cite{random_matrix}; 
beyond leading order, it is more general.
A mean field approximation to the effective
theory gives a random matrix model \cite{random_matrix}.
Such models 
are dominated by eigenvalue repulsion from the Vandermonde determinant
in the measure.  For a $SU(\infty)$ gauge theory in a small volume,
deconfinement is driven by exactly such a mechanism \cite{aharony}.  
I indicate later how eigenvalue repulsion might arise in infinite volume,
from the Higgs effect for an adjoint matrix.

By the converse of asymptotic freedom, 
the running QCD coupling, $\alpha_s(T) = g^2(T)/(4 \pi)$, 
increases as the temperature decreases.  Thus 
a natural possibility is that in the sQGP,
$\alpha_s(T)$ becomes very large as the temperature $T \rightarrow T_c^+$.
For the phenomenology of a strongly coupled,
deconfined phase, see \cite{sQGP}.

A definitive value for $\alpha_s(T)$ can be obtained by matching
correlation functions, for the original theory in four dimensions,
with an effective theory in three dimensions
\cite{braaten,dogma1,andersen,dogma2,dogma3,dogma4}:
\begin{eqnarray}
{\cal L}_{{\it small} \; A_0}^{\it eff}(A_i,A_0)
&=& \; \frac{1}{2} \; {\rm tr}\, G_{i j}^2 
\; + \;{\rm tr}\, \left|D_i A_0\right|^2 \\
+ \; m_D^2 \; {\rm tr}\, A_0^2 &+& \kappa_1 \left({\rm tr}\, A_0^2\right)^2 
\; + \;\kappa_2 \, {\rm tr}\, A_0^4 \; .\nonumber
\label{pert_eff_lag}
\end{eqnarray}
This is the Lagrangian for a massive, adjoint scalar field, $A_0$,
coupled to static magnetic fields, $A_i$:
$A_0$ and $A_i$ are the time like and space like components 
of the vector potential, 
$G_{i j}$ is the non-abelian magnetic field strength, 
and $D_i$ the covariant derivative.
Fields and couplings are normalized
as in four dimensions, with the 
three dimensional action $\int d^3 x/T$ times the Lagrangian.
At leading order, integrating out
the four dimensional modes produces a Debye mass for $A_0$, 
$m_D^2/T^2 \sim \alpha_s$, 
and quartic couplings, $\kappa_1$ and $\kappa_2$, 
$\sim \alpha_s^2$, with each a power series in $\alpha_s$.

This effective theory represents an optimal resummation of perturbation
theory.  As such, it applies only when
fluctuations in $A_0$ are small.  Computing the pressure
to four loop order, $\sim \alpha_s^3$, the 
results are complete up to one undetermined constant
\cite{dogma4}.  Even with the most favorable choice
for this constant, however, the pressure does not agree with that from
numerical simulations on the lattice below temperatures of $\sim 3 T_c$
\cite{dogma1,dogma2}.  

These computations are done in imaginary time, where the ``energies''
are multiples of $2 \pi T$.  Thus 
the coupling constant $\alpha_s(T)$ runs with a scale 
which is of order $\sim 2 \pi T$ \cite{braaten}.  
Computations to two loop order show that even better,
this mass scale is $\sim 9 T$ in QCD \cite{dogma2}.  For
$T_c \sim 175$~MeV, this is $\sim 1.6$~GeV;
at $3 T_c$, it is $\sim 4.7$~GeV.
While these mass scales are not asymptotic, neither are
they obviously in a non-perturbative regime: {\it e.g.},
$\alpha_s(1.6$~GeV$)\sim 0.28$ \cite{dogma2}.  
Hence the question becomes: why does this effective theory fail between
$T_c$ and $\sim 3 T_c$, if the coupling is {\it not} that large?

To see how this might occur, consider a straight, thermal Wilson line
in the fundamental representation:
\vspace{-.05in}
\begin{equation}
\boldl(x,\tau) = {\textit {\large P}} \;
{\textit {\Large e}}^{\; \displaystyle i  g
\displaystyle \int^{\tau}_0 \displaystyle A_0(x,\tau') \; d\tau' } \; ;
\label{def_wilson_line}
\end{equation}
$\it P$ denotes path ordering, $x$ is the spatial
position, and $\tau$, the imaginary time, runs from $0$ to $1/T$.
A closed loop is formed by wrapping all of the way around in imaginary time,
$\boldl(x,1/T)$.  As this quantity arises frequently, 
I denote it by $\boldl(x)$.

The Wilson line is a matrix in color space, and so is not directly
gauge invariant: under a gauge transformation ${\cal U}(x,\tau)$,
$
\boldl(x) \rightarrow {\cal U}^\dagger(x,1/T) 
\boldl(x) \, {\cal U}(x,0)
$.
The trace of the Wilson line is gauge invariant,
and is the Polyakov loop in the fundamental
representation.  Normalizing so that this loop is one when $A_0 = 0$, then
its expectation value should be near one if $g A_0/(2 \pi T)$ is small.
Numerical simulations of a lattice $SU(3)$ gauge theory 
show that while the expectation
value of the renormalized triplet loop is near one at $3 T_c$, 
this is not so when $T < 3 T_c$.
Without dynamical quarks, it drops to a value of
$\approx 0.45$ at $T_c$ \cite{dhlop,dlp,bloop1,adjoint}; its value
with dynamical quarks is similar \cite{bloop2}.

Since the triplet loop is significantly less than
one between $T_c$ and $\sim 3 T_c$, 
in this region it is necessary to extend the 
program of \cite{braaten,dogma1,dogma2,dogma3,dogma4}
to construct an effective, three dimensional theory for
arbitrary values of $g A_0/(2 \pi T)$.
While $A_0$ can be large, as it  
applies only for distances $\gg 1/T$, we can
assume that all spatial momenta are small relative to $2 \pi T$
\cite{gpy,weiss,interface,rdp,diakonov,megias}.   This is
like chiral perturbation theory, with temperature playing the role
of the pion decay constant.

Certainly the effective theory must be invariant 
under static gauge transformations,
${\cal U}(x,\tau)={\cal U}(x)$.  In addition, and somewhat
unexpectedly for a theory in three dimensions,
certain time dependent gauge transformations matter.
For a $SU(N)$ gauge group, consider 
\begin{equation}
{\cal U}_c(\tau) = 
\textit{{\large e}}^{\, 
\textstyle 2 \pi i \, \tau T \, t_N} \; , \;
t_N =
\left(
\begin{array}{cc}
1_{N-1} & 0      \\
0             & -(N-1) \\
\end{array}
\right) \; ;
\label{gauge_transf}
\end{equation}
$1_{N-1}$ is the unit matrix.  
This is spatially constant and strictly periodic in $\tau$,
${\cal U}_c(1/T) = {\it e}^{2 \pi i} \, 1_N = {\cal U}_c(0)$,
and so appears to be rather trivial.
Instead, it turns out to be essential
in constraining the form of the effective Lagrangian at large $A_0$.
Since they don't alter the boundary conditions in imaginary time,
similar gauge transformations exist for any gauge group, 
coupled to matter fields in arbitrary representations.

The problem 
cannot be ignored at large $A_0$, and arose previously
\cite{diakonov,megias}.  The $N^{th}$
root of ${\cal U}_c$ is an aperiodic gauge transformation,
$= {\it e}^{2 \pi i/N} 1_N$ at $\tau = 1/T$.  If there are no
dynamical quarks present,
this is an allowed gauge transformation, and reflects
the $Z(N)$ center symmetry of a $SU(N)$ gauge group \cite{G2}.
Ref. \cite{diakonov} computed 
in the presence of nonzero, background fields for both $A_0$ 
{\it and} $A_i$, allowing $A_0$ to be large.
They found that if the effective Lagrangian is formed from 
terms such as $D_i A_0$, then 
the $Z(N)$ center symmetry appears to be violated
at one loop order.
The argument above, applied to ${\cal U}_c^{1/N}$, shows that even 
classically, $E_i = D_i A_0$ is not consistent
with the requisite $Z(N)$ symmetry.

The significance of these large gauge transformations can be understood by
looking at the Wilson line.  Since $\boldl$ is a $SU(N)$ matrix,
$\boldl^\dagger(x) \boldl(x) = 1_N$, 
it can be diagonalized by a unitary transformation \cite{hooft},
\begin{equation}
\boldl(x) = \Omega(x)^\dagger \, {\it e}^{i \lambda(x)} \,\Omega(x) \; .
\label{diagL}
\end{equation}
$\lambda(x)$ is a diagonal matrix, with elements
$\lambda_a$, $a= 1 \ldots N$.  As $\det(\boldl) = 1$,
${\rm tr} \, \lambda(x) = 0$, modulo $2 \pi$.
Under static gauge transformations, ${\cal U}(x)={\cal U}$,
the adjoint covariant derivative and the Wilson line transform
similarly, 
$D_i \rightarrow {\cal U}^\dagger D_i \, {\cal U}$ and
$\boldl\rightarrow {\cal U}^\dagger \boldl \, {\cal U}$.
Hence the $\lambda_a$ do not change, while $\Omega$ is gauge dependent,
$\Omega \rightarrow \Omega \, {\cal U}$
\cite{hidden}.

The $\lambda_a$
can change under time dependent gauge transformations:
under (\ref{gauge_transf}),
$\lambda \rightarrow \lambda + 2 \pi t_N$, so each
$\lambda_a$ shifts by an integral multiple of $2 \pi$.  Hence gauge
transformations such as (\ref{gauge_transf}) ensure that
the $\lambda_a$'s are periodic
variables.  Of course this is obvious from the definition of
the Wilson line, since its eigenvalues are just ${\it e}^{i \lambda_a}$.

This periodicity is present for an abelian gauge group, where
the Wilson line is merely a phase, $\boldl = {\it e}^{i \lambda}$.
Shifting $\lambda \rightarrow \lambda + 2 \pi$
is an Aharonov-Bohm effect, where the Wilson line, in imaginary time,
wraps around a patch of magnetic flux in a fictitious fifth dimension.
This illustrates elementary topology \cite{math}.
At nonzero temperature, imaginary time is isomorphic to a sphere
in one dimension, $S^1$.  Topologically nontrivial windings are given by
mappings from $S^1$ into $U(1)$, and are classified by
the first homotopy group, $\pi_1(U(1)) = {\cal Z}$, 
where $\cal Z$ is the group of the integers.

The result for a nonabelian group is an exercise in
abelian projection \cite{hooft}.
The $r$ diagonal generators in the Cartan subalgebra
define the maximal torus, which is an
abelian subgroup of $U(1)^r$, the direct product of $r$ $U(1)$'s.
Nontrivial windings are then given by $\pi_1(U(1)^r) = {\cal Z}^r$.
In $SU(N)$, $r=N-1$, and $t_N$ is one of these diagonal generators.

The effective Lagrangian must respect the periodicity of the 
$\lambda_a$'s.  This
is automatic if it is constructed from the Wilson line.  What is then
obscure is the form of the effective electric field, $E_i$.  Consider
\begin{equation}
E_i(x) = \frac{T}{ig} \; \boldl^\dagger(x) D_i \boldl(x) \; .
\label{electric}
\end{equation}
Like the original electric field, this is gauge covariant,
$E_i \rightarrow {\cal U}^\dagger E_i {\cal U}$.  It is also
hermitean, and so is not $ \sim D_i \boldl$.
If the gauge group has a center symmetry, then $E_i$ is trivially
center symmetric.  In accord with the conclusions of \cite{G2}, though,
the presence of a center symmetry is really secondary for what follows.

For small $A_0$, and static $A_i \neq 0$, this reduces to the expected form,
$E_i = D_i A_0$, as in (\ref{pert_eff_lag}).
This rules out using an $E_i$ constructed entirely
from the eigenvalues of $\boldl$.
The simplest example is
$E_i \sim \partial_i \lambda$, with an infinity of other terms,
such as $|{\rm tr} \boldl|^2$ times this, {\it etc.}

There is one last limit which is essential in establishing (\ref{electric}),
although its origin will only be clear after the discussion of interfaces
below.  
I require that when $A_i = 0$, and $A_0$ is static and diagonal ---
but of {\it arbitrary} magnitude --- that it reduces to the abelian form,
$E_i = \partial_i A_0$.  This 
forbids an infinity of terms, formed
by taking various combinations of traces of $\boldl$ times
(\ref{electric}), such as $|{\rm tr} \boldl|^2$, $|{\rm tr} \boldl^2|^2$, 
{\it etc.}  (Equivalently, one can write these terms
times (\ref{electric_action}),
as in (9)-(13) of \cite{oswald}.)  To leading order, these conditions
uniquely determine $E_i$.
In mathematics, (\ref{electric}) is known
as the left invariant one form of $\boldl$ \cite{nair}.

Using the properties of path ordering, the effective electric field 
can be written as
\begin{equation}
E_i/T = \int^{1/T}_0 d\tau \; 
\boldl(\tau)^\dagger \; \partial_i A_0(\tau) \; \boldl(\tau) 
- \boldl^\dagger \left[ A_i , \boldl\right]\; ;
\label{electric_explicit}
\end{equation}
%E_i(x)/T &=& \int^{1/T}_0 d\tau \; 
%\boldl(x,\tau)^\dagger \; \partial_i A_0(x,\tau) \; \boldl(x,\tau) \nonumber \\
% &-& 
%\boldl(x,1/T)^\dagger \left[ A_i(x) , \boldl(x,1/T)\right]\; .
%\label{electric_explicit}
%\end{eqnarray}
%
$\boldl = \boldl(1/T)$, and the $x$ dependence is suppressed.
Up to the various Wilson lines --- which are, after all, 
elements in the gauge group --- this is
a plausible form for a gauge covariant 
electric field formed by averaging over $\tau$.

With this $E_i$, the effective Lagrangian is
that of a gauged,
nonlinear sigma model \cite{random_matrix,strong}:
\begin{equation}
{\cal L}^{\it eff}_{\it classical}(A_i,\boldl) = 
\; \frac{1}{2} \; {\rm tr}\; G_{i j}^2 + \frac{T^2}{g^2} 
\; {\rm tr}\, \left|\boldl^\dagger D_i \boldl\right|^2 \; .
\label{classical}
\end{equation}
Using the decomposition of the Wilson line in 
(\ref{diagL}), the electric field term is proportional to
\begin{equation}
{\rm tr}\, \left|D_i \boldl \right|^2 \; = \;
{\rm tr}\, \left(\partial_i \lambda\right)^2 \;
+ \; {\rm tr}\, 
\left|\left[\Omega \, D_i \, \Omega^\dagger,{\it e}^{i \lambda}
\right]\right|^2 \; .
\label{electric_action}
\end{equation}
The first term is that of an abelian theory, while the second
couples the nonabelian electric and magnetic sectors together.
Since ${\it e}^{i \lambda}$ is invariant under static
gauge transformations, so is the combination
$\Omega \, D_i \, \Omega^\dagger$ \cite{hidden}.

On the lattice, the analogy of 
(\ref{electric_action}) is well known from Banks and Ukawa \cite{banks_ukawa}.
I suggested (\ref{electric_action}) previously \cite{rdp}, but only
by expanding in small $A_0$.  This does not suffice to fix
its form at large $A_0$.  For a related linear model, see \cite{vuorinen}.

The effective Lagrangian of (\ref{classical}) is not renormalizable
in three dimensions (it is in two \cite{oswald}),
but this is a standard feature of effective theories \cite{braaten}.
It is also common that the effective fields are only
indirectly related to those in the original theory, 
although it is especially striking here.

As an aside, I remark that the instanton
number in four dimensions carries over directly to the effective
theory.  Start with a smooth,
strictly periodic classical field, $A_\mu(x,\tau)$,
and then transform to $A_0 = 0$ gauge.  The gauge transformation which
does this is just $\boldl(x,\tau)$, (\ref{def_wilson_line}).
The instanton number is then a difference of Chern-Simons terms between
$\tau = 1/T$ and $0$ \cite{gpy}.  One can show that the instanton
number equals the winding number of the Wilson line:
\begin{equation}
\frac{1}{24 \pi^2} \; \int d^3 x \;\epsilon^{i j k} \;
{\rm tr}\, ( C_i C_j C_k ) \;\;\; ; \;\;\;
C_i = \boldl^\dagger \partial_i \boldl \; ,
\end{equation}
which is an integer.  This suggests an analogy to the
color Skyrmions of \cite{nair}: these have nonzero winding number, but
are not instantons.

To establish (\ref{classical}), it is necessary to show that 
it gives the same physics as the original theory,
especially at large $A_0$.  
One possibility is to use the interfaces
which exist because the $\lambda_a$'s are periodic.

This is most familiar for the $Z(N)$ interface of a $SU(N)$ gauge theory
without quarks \cite{interface,vuorinen}.  
This is given by taking a box, long in one spatial direction, with
$\boldl = 1_N$ at one end of the box, and
$\boldl = {\it e}^{2 \pi i/N} 1_N$ at the other.   
A $Z(N)$ interface is related to the disorder parameter of 
't Hooft \cite{kovner,forcrand}.  

The interface which corresponds to ${\cal U}_c$,
(\ref{gauge_transf}), is given 
by taking $A_0 = 0$ at one end of the box, and $A_0 = 2 \pi T t_N/g$ at
the other \cite{digiacomo}.  This is a $U(1)$ interface:
while $\boldl = 1_N$ at both ends of the box,
the change in $A_0$ is nontrivial, and cannot be undone \cite{math}.
It is not related to a disorder parameter.
Without quarks, there is 
a row of $N$, distinct $Z(N)$ interfaces.  With quarks,
these coalesce into one $U(1)$ interface;
like the expectation value of $\boldl$, it exists in both phases. 

To leading order in $g^2$, it is easy to use interfaces
to match the effective theory to the original.
In the original theory, compute for constant $\boldl$
to one loop order.  For a $SU(N)$ gauge theory without
quarks, this gives \cite{gpy}
\begin{equation}
{\cal L}^{\it eff}_{1\; loop}(\boldl)
= \; - \; \frac{2 \,T^4}{\pi^2} \; \sum_{m=1}^{\infty}
\; \frac{1}{m^4} \;
\left| {\rm tr}\, \boldl^m \right|^2 + 
\frac{\pi^2 T^4}{45} \; .
\label{one_loop}
\end{equation}
To leading order, the effective Lagrangian is the sum of
(\ref{classical}) and (\ref{one_loop}) \cite{interface}.
Because $E_i = \partial_i A_0$ when $A_i = 0$,
and $A_0$ is static, diagonal, and of arbitrary magnitude,
trivially a $Z(N)$ interface is the same in both theories.
Dynamical quarks add new terms to the potential,
which lift the $Z(N)$ symmetry, and so remove $Z(N)$ interfaces.
$U(1)$ interfaces remain, 
and are analyzed similarly, with the same result for $E_i$.

At higher order, matching between the original and effective theories
is much more involved.  The effective Lagrangian is constructed
from $\boldl$ and $G_{i j}$ in a derivative expansion, with
terms for constant $\boldl$ \cite{aharony,dhlop,dlp},
two derivatives \cite{dhlop,interface,rdp,diakonov,megias,oswald},
four \cite{diakonov}, and so on.  
At higher order, matching will involve computing both the interface
tension, $\sim T^2/\sqrt{\alpha_s}$, 
and expectation values of gauge invariant operators in the presence
of an interface.  

Further, my discussion of (\ref{pert_eff_lag}) was incomplete: it is
merely the first step of three,
with the others integrating out the electric and magnetic
sectors \cite{braaten,dogma1,dogma2,dogma3,dogma4}.
While $A_0$ is large at the center of
an interface, it is small at the ends, and so there the electric
sector must be treated more carefully.
For the pressure, it should be possible to isolate that piece which
is $\boldl$ dependent, after subtracting the vacuum energy
of the static magnetic sector for $\boldl = 0$ \cite{dogma4}.

While meaningful statements can only be made after computation at
next to leading order, when the scale of $\alpha_s$ is set,
qualitatively
much of the physics can be understood from (\ref{one_loop}).
The perturbative vacuum, $\langle \boldl \rangle = 1_N$, 
gives minus the pressure of an ideal $SU(N)$ gas,
$
p_{\it ideal} = - 
{\cal L}^{\it eff}_{1 \; {\it loop}}(1_N) =
+ (N^2-1 ) \pi^2 T^4/45
$.
This is the absolute minimum at leading order, and it is
at least metastable, order by order in $\alpha_s$.

In a $SU(N)$ gauge theory without quarks, deconfinement is related
to the breaking of a global $Z(N)$ symmetry:
under a $Z(N)$ transformation, $\boldl \rightarrow z \boldl$,
where $z = {\it e}^{2 \pi i/N}$.
Consider the diagonal $SU(N)$ matrix
\begin{equation}
\boldl_c = {\rm diag}\left(1,z,z^2\ldots z^{N-1} \right) \;
\label{boldlc}
\end{equation}
Of the loops constructed from $\boldl_c$, only those which are $Z(N)$
neutral are nonzero: if $m$ is an integer, 
${\rm tr}\, (\boldl_c)^{m} = 0$ when $m$ is not a multiple of $N$, and
$= N$ when it is.
Hence $\boldl_c$ might represent the $Z(N)$ symmetric, confined vacuum 
\cite{weiss,polchinski,schaden}.
However, at leading order, (\ref{one_loop}), 
$
p(\boldl_c) = - {\cal L}^{\it eff}_{1 \; {\it loop}}(\boldl_c) =
- ( 1 - 1/N^2 ) 
\pi^2 T^4/45 
$.
Thus for any finite $N$, $\boldl_c$ 
has negative pressure, and is not a physical state.

At infinite $N$, however, $\boldl_c$ {\it does} represent the confined vacuum.
While its pressure is negative, this is $\sim 1$, and is 
negligible relative to that $\sim N^2$ in the deconfined phase
\cite{aharony,polchinski,schaden}.  
While (\ref{one_loop}) is only valid at leading order,
since any trace of $\boldl_c$ vanishes at $N=\infty$,
the pressure for $\boldl_c$ remains $\sim 1$ to all orders
in $\alpha_s N$ \cite{schaden}.

At infinite $N$, $\boldl_c$ is familiar from random matrix models:
there is complete eigenvalue repulsion, and a flat eigenvalue density
\cite{random_matrix,aharony,polchinski}.  
Numerical simulations suggest that in the confined phase,
the eigenvalue density for small $N$ is like that of $N=\infty$.
By factorization, in the confined phase the expectation
value of the renormalized adjoint loop is $\sim 1/N^2$
\cite{dhlop}.  For $N=3$, though, numerically this is found to be not
$\sim 10\%$, but only $\sim 1\%$ \cite{dhlop,adjoint}.  
That the expectation value of a $Z(N)$ neutral loop is so small indicates that
the functional integral is close to an integral over the 
group measure; {\it i.e.}, that the eigenvalue density is nearly flat.

In perturbation theory, though, there is no sign of any eigenvalue
repulsion which might produce a flat distribution.
As in (\ref{one_loop}), and seen to three loop order in \cite{aharony},
the perturbative potential for 
constant $\boldl$ only involves sums of eigenvalues, and not
differences.  Thus eigenvalue repulsion, and so confinement,
must be generated by fluctuations in the effective theory.

It is known how this happens for $SU(\infty)$ on
a very small sphere \cite{aharony}.
The effective Lagrangian is a single integral
for the constant mode of $\boldl$: as
a random matrix model, 
the Vandermonde determinant in the measure
generates eigenvalue repulsion and drives the transition 
\cite{random_matrix,aharony,dlp}.
In infinite volume, though,
terms in the measure depend upon the regularization;
{\it e.g.}, they vanish with dimensional regularization.

To represent the non-perturbative effects which might drive
the transition in infinite volume, consider adding to the effective Lagrangian 
\begin{equation}
{\cal L}^{\it eff}_{non-pert.}
\sim + B_f \, T^2 \, |{\rm tr}\, \boldl |^2 \; .
\label{bf}
\end{equation}
While motivated by precise results from numerical simulations 
\cite{bielefeld}, this term is only meant to illustrate
what is possible near $T_c$ \cite{fuzzy}.  It 
shifts the minimum in the loop potential 
from the perturbative value,
$\langle \boldl \rangle = 1_N$, to some
$\langle \boldl \rangle \neq 1_N$.  This is interesting because it
produces a Higgs effect for $A_i$.  
As an adjoint field, in perturbation theory
the mass magnetic gluons acquire from
$\langle \boldl \rangle \neq 1_N$ involves
differences of eigenvalues, as diagonal gluons remain massless, 
and off diagonal gluons develop a mass \cite{hooft}.
Integrating out fluctuations in $A_i$ and $\boldl$ to one loop order
(which is easiest in unitary gauge), there is 
a qualitatively new term in the effective Lagrangian,
\begin{equation}
\Delta {\cal L}^{\it eff} \sim - \sum_{a,b=1}^N
\left( g^2 |{\it e}^{i \lambda_a} - {\it e}^{i \lambda_b}|^2
\right)^{3/2} \; .
\label{new}
\end{equation}
The mass dimensions are made up by $T$ and $B_f$.
The sign is physical, and corresponds to eigenvalue repulsion.
Once (\ref{new}) is included, 
$\langle \boldl \rangle$ is given by 
a distribution of eigenvalues.

These calculations are only suggestive.  It is not
obvious how to characterize, gauge invariantly,
such an adjoint Higgs phase for strongly
coupled gluons in three dimensions.
Qualitatively, a Higgs phase should increase the mixing between
the Wilson line and magnetic glueballs \cite{vuorinen,higgs},
which is usually very small.

The effective theory, as determined perturbatively,
can be studied in various ways.
Since the ultraviolet cutoff is physical, it is reasonable to start
with mean field theory \cite{dhlop,dlp}.  To do better, the
theory could be simulated numerically on a lattice, to 
directly measure quantities
such as the eigenvalue distribution, glueball masses, {\it etc}. 
At large $N$ \cite{teper}, analytic 
approximations may help \cite{yaffe}.

The usual justification for an effective Lagrangian is the
presence of a small mass scale, but generically, there is none
here.  If the effective
coupling is small at $T_c$ \cite{dogma2}, though, then with care
nothing is lost by going to an effective Lagrangian.  
Presumably, this is worthwhile when (\ref{pert_eff_lag}) fails:
from $\sim 3 T_c$, down to some point below $T_c$ \cite{low}.
For constant $\boldl$, the effective potential 
shows no signs of a transition to a confining phase:
at $N=\infty$, and perhaps even for small $N$, this
must involve eigenvalue repulsion.  In infinite volume,
this arises dynamically, especially 
from fluctuations in the angular variables, $\Omega$, and 
the gauge fields, $A_i$.
This, then, is why the effective theory is of interest: we can
use it to uniquely
isolate the dynamic origin of the transition, as eigenvalue
repulsion.  It thus provides a notable example of a
field theory of (not so) random matrices \cite{random_matrix}.

Acknowledgements: This research was supported by D.O.E.
grant DE-AC02-98CH10886, and in part by
the Alexander von Humboldt Foundation.
I gratefully acknowledge discussions with D. Diakonov, who,
during a sabbatical which I took at the Niels Bohr Institute
in '03-'04,
stressed to me that the usual effective Lagrangians are not center symmetric.
I am especially indebted to V. P. Nair and 
C. P. Korthals Altes for discussions; also, to
D. Boer, J. Berges, M. Creutz, O. Kaczmarek, F. Karsch, 
D. Kharzeev, M. Laine, A. Vuorinen, and L. Yaffe for their comments.


\begin{thebibliography}{999}
\bibitem{higgs}
To see changes in mixing, it may be useful
to measure magnetic masses from
spatial plaquettes split in $\tau$: take two spatial links at
$\tau =0$, and two at $\tau = 1/T$, tied together by two thermal Wilson lines.
\bibitem{low}
At low temperature, dimensional reduction is not of use,
and it cannot suffice to include only
straight, thermal Wilson lines; those which
oscillate in time also contribute.  At zero temperature, for
instance, only closed loops matter.
\end{thebibliography}
\end{document}